\documentclass[fleqn]{article}
\usepackage{amsfonts}

\def\noi{\noindent}

\makeatletter
\renewcommand{\section}{\@startsection{section}{1}{0pt}%
        {-3.5ex plus -1ex minus -.2ex}{2.3ex plus .2ex}%
        {\large\bf\protect\raggedright}}

\renewcommand{\subsection}{\@startsection{subsection}{2}{0pt}%
        {-3ex plus -1ex minus -.2ex}{1.4ex plus .2ex}%
        {\normalsize\bf\protect\raggedright}}

\renewcommand{\thesubsubsection}%
        {\arabic{section}.\arabic{subsection}.\arabic{subsubsection}.}

\renewcommand{\@oddhead}{\raisebox{0pt}[\headheight][0pt]{%
   \vbox{\hbox to\textwidth{\rightmark \hfil \rm \thepage \strut}\hrule}}}
\renewcommand{\@evenhead}{\raisebox{0pt}[\headheight][0pt]{%
   \vbox{\hbox to\textwidth{\thepage \hfil \leftmark \strut}\hrule}}}
\newcommand{\heads}[2]{\markboth{\protect\small\it #1}{\protect\small\it #2}}

\makeatother

\def\prepno#1#2
    {\thispagestyle{empty}
    \noi    \unitlength=1mm
    	\begin{picture}(178,10)
            \put(177,15){\llap{\bf #1}}
            \put(177,10){\llap{\small\rm #2}}
        \end{picture}
          }             

\newcommand{\Title}[1]{\noi {\uppercase{\Large #1}} \\}

\newcommand{\Authors}[4]{\noi
        {\large\bf #1\dag\ #2\ddag}\medskip\begin{description}
        \item[\dag]{\it #3} \item[\ddag]{\it #4}\end{description}}

\newcommand{\Abstract}[1]{\vskip 2mm \begin{center}
        \parbox{16.4cm}{\small\noi #1} \end{center}\medskip}

\newcommand{\Theorem}[2]{\medskip\noi {\bf #1. \ }{\sl #2}\medskip}

\newcommand{\sect}[1]{Sec.\,#1}

\def\nqq{\hspace*{-2em}}
\def\nhq{\hspace*{-0.5em}}

\def\cm{\hspace*{1cm}}

\def\Jl#1#2{{\it #1\/} {\bf #2},\ }

\def\ApJ#1 {\Jl{Astroph. J.}{#1}}
\def\CQG#1 {\Jl{Class. Quantum Grav.}{#1}}
\def\DAN#1 {\Jl{Dokl. AN SSSR}{#1}}
\def\GC#1 {\Jl{Grav. \& Cosmol.}{#1}}
\def\GRG#1 {\Jl{Gen. Rel. Grav.}{#1}}
\def\JETF#1 {\Jl{Zh. Eksp. Teor. Fiz.}{#1}}
\def\JETP#1 {\Jl{Sov. Phys. JETP}{#1}}
\def\JHEP#1 {\Jl{JHEP}{#1}}
\def\JMP#1 {\Jl{J. Math. Phys.}{#1}}
\def\NPB#1 {\Jl{Nucl. Phys.}{B\ #1}}
\def\NP#1 {\Jl{Nucl. Phys.}{#1}}
\def\PLA#1 {\Jl{Phys. Lett.}{#1A}}
\def\PLB#1 {\Jl{Phys. Lett.}{#1B}}
\def\PRD#1 {\Jl{Phys. Rev.}{D\ #1}}
\def\PRL#1 {\Jl{Phys. Rev. Lett.}{#1}}

\def\al{&\nhq}
\def\lal{&&\nqq {}}
\def\eq{Eq.\,}
\def\eqs{Eqs.\,}
\def\beq{\begin{equation}}
\def\eeq{\end{equation}}
\def\bear{\begin{eqnarray}}
\def\bearr{\begin{eqnarray} \lal}
\def\ear{\end{eqnarray}}
\def\earn{\nonumber \end{eqnarray}}
\def\nn{\nonumber\\ {}}
\def\nnv{\nonumber\\[5pt] {}}

\def\yy{\\[5pt] {}}
\def\yyy{\\[5pt] \lal }
\def\eql{\al =\al}
\def\sequ#1{\setcounter{equation}{#1}}

\def\dst{\displaystyle}
\def\tst{\textstyle}
\def\fracd#1#2{{\dst\frac{#1}{#2}}}
\def\fract#1#2{{\tst\frac{#1}{#2}}}
\def\Half{{\fracd{1}{2}}}
\def\half{{\fract{1}{2}}}

\def\e{{\,\rm e}}
\def\d{\partial}

\def\const{{\rm const}}

\def\then{\ \Rightarrow\ }

\newcommand{\lims}[1]{\mathop{#1}\limits}
\newcommand{\limr}[4]{\,\raisebox{#1}{${\lims{#2}^{#3}_{#4}}$}\,}

\def\ssph{static, spherically symmetric}

\def\mn{_{\mu\nu}}
\def\MN{^{\mu\nu}}
\def\mN{_\mu^\nu}

\def\R{{\mathbb R}}

\def\cA{{\cal A}}

\def\cK{{\cal K}}
\def\tT{{\widetilde T}}
\def\Te{\limr {1pt} {T}{}{\rm e}}
\def\Ts{\limr {1pt} {T}{}{\rm s}}

\def\thd{\fract 13}

\def\sumi{\sum_{i=1}^3}
\def\ui{{\underline{i}}}

\def\ha {\hat {a}}
\def\hg {\hat {g}}
\def\kappa{\mbox{\sl \ae}}

\heads
{K.A. Bronnikov, E.N. Chudaeva and G.N. Shikin}
{Magneto-dilatonic Bianchi-I cosmology:
	isotropization and singularity problems}

\begin{document}
\prepno{gr-qc/0401125}{}

\Title 	{Magneto-dilatonic Bianchi-I cosmology:\yy
	isotropization and singularity problems}

\Authors{K.A. Bronnikov}{E.N. Chudaeva and G.N. Shikin}
	{VNIIMS, 3-1 M. Ulyanovoy St., Moscow 119313, Russia;\\
    	 Institute of Gravitation and Cosmology, PFUR,
         6 Miklukho-Maklaya St., Moscow 117198, Russia}
	{Department of Theoretical Physics, PFUR,
         6 Miklukho-Maklaya St., Moscow 117198, Russia}

\Abstract{We study the evolution of Bianchi-I space-times filled with a
	global unidirectional electromagnetic field $F\mn$ interacting with
	a massless scalar dilatonic field according to the law $\Psi(\phi)
	F\MN F\mn$ where $\Psi(\phi) > 0$ is an arbitrary function. A
	qualitative study, among other results, shows that (i) the volume
	factor always evolves monotonically, (ii) there exist models
	becoming isotropic at late times and (iii) the expansion generically
	starts from a singularity but there can be special models starting
	from a Killing horizon preceded by a static stage. All three
	features are confirmed for exact solutions found for the usually
	considered case $\Psi = \e^{2\lambda\phi}$, $\lambda = \const$.
	In particular, isotropizing models are found for $|\lambda| >
	1/\sqrt{3}$. In the special case $|\lambda| = 1$, which corresponds
	to models of string origin, the string metric behaviour is studied
	and shown to be qualitatively similar to that of the Einstein frame
	metric.
	}

\section{Introduction}

    The influence of intergalactic magnetic fields on the cosmological
    evolution has been studied for over four decades from both theoretical
    and observational points of view \cite{rev-magn}. Cosmologists speculate
    that such a field could be primordial in origin, i.e., could have
    appeared before nucleosynthesis and even before inflation.

    A cosmological model which contains a global magnetic field is
    necessarily anisotropic since the magnetic field vector specifies a
    preferred spatial direction. The presently observed Universe is almost
    isotropic at large, therefore the isotropization problem appears
    inevitably in any study of anisotropic cosmologies. The simplest of
    such models, which nevertheless rather completely describe the
    anisotropy effects, are Bianchi type I homogeneous models whose spatial
    sections are flat but the expansion or contraction rate is
    direction-dependent.

    A number of recent papers [2--4] 		
    discussed the properties of Bianchi-I cosmologies with global magnetic
    fields in the framework of dilaton gravity with the action (in the
    Einstein frame)
\beq
	S = \int \sqrt{g}\, d^4x \Bigl[ R + 2 (\d\phi)^2         \label{dil}
			- \e^{2\lambda\phi} F\MN F\mn \Bigr],
\eeq
    where $g = |\det g\mn|$, $R$ is the scalar curvature, $\varphi$ is the
    scalar dilaton, $F\mn = \d_\mu A_\nu - \d_\nu A_\mu$ is
    the electromagnetic field, $(\d\varphi)^2 := g\MN \d_\mu\varphi\,
    \d_\nu\varphi$, and $\lambda$ is a coupling constant\footnote
{Our sign conventions are: the metric signature $(+{}-{}-{}-)$, the curvature
 tensor $R^{\sigma}_{\ \mu\rho\nu} = \d_\nu\Gamma^{\sigma}_{\mu\rho}-\ldots$,
 so that, e.g., the Ricci scalar $R > 0$ for de Sitter space-time, and the
 stress-energy tensor (SET) such that $T^t_t$ is the energy density.
 The gravitational constant is absorbed by re-definition of the fields
 $\phi$ and $F\mn$.}.
    This theory with $\lambda = \pm 1$ naturally appears in the weak field
    limit of string theories as part of their bosonic sector and is
    widely discussed in cosmological problems. Thus, there are attempts to
    apply dilatonic anisotropic cosmologies for describing some stages of
    the string-motivated pre-big-bang scenario \cite{gio99,gio00}.

    Refs. \cite{salim}--\cite{harko} contain exact solutions to the
    Einstein-Maxwell-dilaton equations due to (\ref{dil}), obtained in
    different variables. A general conclusion of these papers is that the
    evolution of expanding Bianchi-I models starts from a singularity of
    zero volume and ends with an infinite expansion, and that isotropization
    can never be achieved in this class of models. For the ``stringy''
    coupling $\lambda = \pm 1$, it was found \cite{gio99} that isotropy
    could be achieved due to higher curvature corrections to the equations
    of gravity.

    Our purpose in this work was to re-examine the isotropization and
    singularity problems for magneto-dilatonic Bianchi I cosmologies in a
    more general context, invoking an arbitrary coupling function
    $\Psi(\phi)$ [see \eq (\ref{S-Psi})] instead of an exponential function.
    This is, above all, motivated by the considerations of Damour and
    Polyakov \cite{dampol} who have argued that the string loop expansion is
    able to produce a coupling function of the form
    $B(\Phi) = \e^{-2\Phi} + c_0 + c_1 \e^{2\Phi} + \ldots$ ($c_i = \const$)
    instead of simply $\e^{-2\Phi}$ which corresponds to the tree-level
    contribution only.

    It is certainly impossible to obtain a general solution for any given
    $\Psi(\phi)$ but some essential qualitative inferences are available. We
    have been able to find the following:

\begin{enumerate}
\item
    The volume factor $v$ is always a monotonic function of time, so that a
    model can be either permanently expanding or permanently contracting.
\item
    If collinear electric and magnetic fields are simultaneously admitted,
    isotropization is impossible. The further consideration concerns
    purely magnetic models; their electric counterparts can be obtained by
    duality.
\item
    Under some restriction on the choice of $\Psi (\phi)$, it is possible to
    obtain expanding cosmologies with late-time isotropization, and this
    is true for some models with $\Psi \sim \exp (\const\cdot\phi)$.
\item
    The expansion begins (or the contraction ends) either at a singularity
    or at a (Killing) horizon beyond which there is a static space-time
    region.
\item
    Both isotropization and a horizon in the past are only possible for
    plane-symmetric models, in which, among the three scale factors
    $a_i(t)$, two coincide. Special models, which start their expansion from
    a horizon and end with isotropy, are not excluded.
\end{enumerate}

    These results are partly at variance with those of Refs.\,[2--4]
    obtained for the particular action (\ref{dil}). It therefore appears
    reasonable to reconsider this case comparing the results with items
    1--5. We do this and confirm our general observations. Moreover, we give
    explicit conditions under which the model is asymptotically isotropic at
    late times and at which there is a horizon at the beginning. These
    conditions coincide only partly, but, for the special value of the
    coupling constant $\lambda = \sqrt{3}$, there exists a two-parameter
    family of solutions combining these two features. We also show that the
    static region beyond the horizon (in the absolute past with respect to
    the cosmological evolution) contains a Reissner-Nordstr\"om-like
    singularity.

    The paper is organized as follows. \sect 2 presents the field equations
    for an arbitrary coupling function $\Psi (\phi)$, written in terms of
    the harmonic time coordinate $u$ which considerably simplifies the
    treatment. \sect 3 contains a qualitative study of Bianchi-I cosmologies
    with this arbitrary coupling function. This study is preceded by a short
    discussion of isotropization criteria. In \sect 4 we put $\Psi =
    \e^{2\lambda\phi}$, obtain the full set of exact solutions and briefly
    discuss their properties. We show, in particular, that asymptotically
    isotropic solutions do exist for $\lambda > 1/\sqrt{3}$ and form a
    two-parameter family. In particular, models of string origin ($\lambda =
    1$) can isotropize at late times. Models which expand from a horizon
    rather than a singularity are also considered. In \sect 5 we put
    $\lambda=1$ and briefly describe the properties of the string metric
    $\hg\mn$, which turn out to be quite similar to those of the Einstein
    metric $g\mn$. \sect 6 contains concluding remarks, in particular, on
    exact solutions for the system (\ref{dil}) other than those discussed in
    other sections. Lastly, in the Appendix we give an expression for the
    Kretschmann scalar for Bianchi-I cosmologies needed to check their
    regularity.

\section{Field equations}

    Consider dilaton gravity with an action more general that (\ref{dil}),
    containing an arbitrary coupling function $\Psi(\phi)$:
\beq
     S = \int \sqrt{g} d^4x \Bigl[ R + 2 (\d\phi)^2          \label{S-Psi}
				- \Psi(\phi) F\MN F\mn \Bigr],
     \cm \Psi(0) = 1,
\eeq
    which leads to the field equations
\bearr
	\frac{1}{\sqrt{g}}\d_\mu\bigl(\sqrt{g}g\MN \d_\nu\phi\bigr)
		   + \frac{1}{4} \Psi_\phi F\MN F\mn =0,     \label{eq-fi}
\\ \lal
	\d_\mu (\sqrt{g}F\MN\Psi(\phi)) =0,                  \label{eq-F}
\yyy
	R\mN = - (T\mN - \half T\delta\mN) \equiv -\tT\mN,    \label{EE}
\ear
    where $T = T_\alpha^\alpha$, and the stress-energy tensor (SET) $T\mN$
    is a sum of the scalar part $\Ts\mN$ and the electromagnetic part
    $\Te\mN$ containing the interaction factor $\Psi(\phi)$:
\bear
      T\mN \eql \Ts\mN + \Te\mN,                            \label{SET}
\nn
      \Ts\mN \eql 2\phi_{,\mu}\phi^{,\nu}
      		- \delta\mN \phi_{,\alpha}\phi^{,\alpha},
\nn
      \Te\mN \eql \Psi(\phi)[-2F_{\mu\alpha}F^{\nu\alpha} +
      			\half \delta\mN F_{\alpha\beta}F^{\alpha\beta}].
\ear

     Now consider the field system (\ref{S-Psi}) in a
     Bianchi type I space-time described by the metric
\bear                                                          \label{B1}
     ds^2 = \e^{2\alpha} dt^2 - \sumi \e^{2\beta_i} (dx^i)^2,
\ear
     where $\alpha$ and $\beta_i$ are functions of the time coordinate $t$.
     The proper physical time of a comoving observer, $t=\tau$, corresponds
     to the coordinate condition $\alpha \equiv 0$. The harmonic time
     coordinate, $t=u$, is obtained if the lapse function $\e^{2\alpha}$ is
     chosen so that
\beq
     \alpha = \beta_1 + \beta_2 + \beta_3.                     \label{harm}
\eeq
     This choice significantly simplifies the Einstein equations. In terms of
     $u$, the nonzero components of the Ricci tensor are
\bear
       R_0^0 = \e^{-2\alpha} \biggl(\ddot{\alpha} -\dot{\alpha}{}^2
		     + \sumi \dot{\beta_i}^2\biggr),
\cm
       R_i^\ui = \e^{-2\alpha} \ddot{\beta}_i.                 \label{R_mn}
\ear
     There is no summing over an underlined index, and the dot denotes
     $d/du$.

     We assume that $\phi = \phi(u)$ and that there are homogeneous
     electric and magnetic fields having the same direction $x^1$. So the
     electromagnetic vector potential is taken in the form
\[
	A_\mu = \{ 0,\ A_1(u),\ 0, A_3(x^2) \}
\]
     Then \eqs (\ref{eq-F}) and the corresponding Bianchi identities lead to
\beq
	F_{01} = - F_{10} = \frac{q_e}{\Psi(\phi)\sqrt{g}}
				       \e^{2\alpha + 2\beta_1},
\cm
	F_{23} = - F_{32} = q_m,
\eeq
     where $q_e$ and $q_m$ are constants characterizing the electric and
     magnetic field intensities, respectively; other $F\mn$ are zero.
     Accordingly, the tensor $\Te\mN$ has the form
\beq                                                            \label{T_e}
      \Te_0^0 = \Te_1^1 = -\Te_2^2 = -\Te_3^3 = (B^2+E^2)\Psi(\phi),
\eeq
     where $E$ and $B$ are the electric and magnetic field strengths:
\beq
      E^2 = F_{01}F^{10} = (q_e^2/\Psi^2)\e^{-2\beta_2 - 2\beta_3},
\cm                                                             \label{E,B}
      B^2 = F_{23}F^{23} = q_m^2 \e^{-2\beta_2 - 2\beta_3}.
\eeq
     One may notice the absence of the usual electric-magnetic duality as a
     result of the interaction. There is still a duality involving $\Psi$:
     the Einstein-scalar equations are invariant under the substitution
     $B^2 \Psi \leftrightarrow E^2/\Psi$.

     The scalar SET is characterized by the only nonzero component of
     ${\tT}\mN$, namely, $\tilde{\Ts}_0^0 = \e^{-2\alpha} \dot{\phi}{}^2$.
     Thus the total SET possesses the symmetry $\tT_1^1 = -\tT_2^2 =
     -\tT_3^3$. Comparing, according to (\ref{EE}), the corresponding Ricci
     tensor components in terms of the harmonic time coordinate
     (\ref{harm}), we obtain
\[
	\ddot \beta_2 = \ddot \beta_3 = -\ddot \beta_1,
\]
     whence it follows
\bear
	\beta_2 \eql -\beta_1 + c_2 u,   \nn
	\beta_3 \eql -\beta_1 + c_3 u,   \nn
	\alpha  \eql -\beta_1 + (c_2+c_3)u,                 \label{sol_23}
\ear
     where $c_2,\ c_3 = \const$ and two more integration constants have been
     removed by constant rescalings of the $x^2$ and $x^3$ axes.

     The remaining unknowns are $\beta_1(u)$ and $\phi (u)$. They can be
     found from \eq (\ref{eq-fi}) and the so far unused components of the
     Einstein equations (\ref{EE}):
\bearr
     \ddot \phi = -\Half \e^{2\beta}                         \label{dd-phi}
     			\Bigl (q_m^2 \Psi_\phi
			     - q_e^2 \Psi_\phi/\Psi^2 \Bigr);
\\ \lal
     \ddot \beta = - \e^{2\beta}           		    \label{dd-beta}
     			\Bigl (q_m^2 \Psi + q_e^2/\Psi \Bigr);
\\ \lal                                                         \label{int}
     \dot\beta{}^2 + \dot\phi{}^2
           + \e^{2\beta} \Bigl( q_m^2 \Psi + q_e^2/\Psi \Bigr) = c_2 c_3,
\ear
     where $\beta \equiv \beta_1$. \eq (\ref{dd-beta}) is a sum of spatial
     components of (\ref{EE}) while (\ref{int}) is the constraint equation
     $\kappa T_0^0 = \ldots$, representing a first integral of
     (\ref{dd-phi}) and (\ref{dd-beta}).

     It is hard to solve these equations for a given function
     $\Psi(\phi)$ unless it is chosen in some special forms, such as,
     e.g., $\Psi = \e^{2\lambda\phi}$ in (\ref{dil}). Before obtaining a
     solution for this, most frequently discussed form of dilaton gravity,
     we would like to discuss the isotropization and singularity problems
     without specifying the function $\Psi(\phi)$. We only assume it to be
     smooth and positive, thus preserving the correct sign of field energy.

\section {Arbitrary $\Psi(\phi)$: isotropization and regularity}

\subsection{Isotropization criteria}

     Isotropization means, by definition, that at large physical times
     $\tau$, when the volume factor $v= v(\tau) = a_1 a_2 a_3 =
     \e^{\beta_1+\beta_2+\beta_3}$ tends to infinity, the three scale
     factors $a_i(\tau) = \e^{\beta_i}$ grow at the same rate, i.e., that
     $a_i/a \to \const$, where $a(\tau) = v^{1/3}$ is the average scale
     factor. In terms of the harmonic time $u$, we have $v =\e^{\alpha(u)}$,
     and, due to (\ref{sol_23}) and (\ref{int}), all the derivatives
     $\dot\alpha$ and $\dot\beta_i$ are restricted above. Therefore $v \to
     \infty$ can only correspond to $u \to \pm \infty$. Choosing $+\infty$,
     we can write the isotropization condition in the form
\beq                                                         \label{iso}
     \beta_i - \beta_k \to \const \cm {\rm as}\quad u \to \infty.
\eeq

     As a measure of anisotropy, one sometimes uses such quantities as the
     mean anisotropy parameter $\cA$ and the shear parameter $\Sigma^2$
     defined as (see, e.g., \cite{harko})
\beq
     \cA = \frac{1}{3}\sumi \frac{H_i^2}{H^2} -1,
\cm
     \Sigma^2 = \frac{3}{2} \cA H^2,     \label{def_Sig}
\eeq
     where $H_i = a_i^{-1} da_i/d\tau = \e^{-\alpha}\dot \beta_i$ are the
     ``directional'' Hubble parameters and $H = a^{-1} da/d\tau
     = \thd \e^{-\alpha} \dot\alpha$ is the mean Hubble parameter. In our
     variables,
\beq                                                          \label{Sig}
     \cA = \frac{3}{\dot\alpha^2}
	   \bigl(\dot\beta_1^2 +\dot{\beta}_2^2+\dot{\beta}_3^2 \bigr) - 1,
\cm
     \Sigma^2 = \frac{1}{6} \e^{-2\alpha}
     	\Bigl[ 3\bigl(\dot\beta_1^2 +\dot{\beta}_2^2+\dot{\beta}_3^2 \bigr)
			- \dot\alpha^2 \Bigr].
\eeq
     The requirement (\ref{iso}) automatically leads to both $\cA
     \to 0$ and $\Sigma^2 \to 0$ as $u\to \infty$, but the converse is
     in general not true.

     Thus, to obtain $\cA \to 0$ it is sufficient to suppose that the
     differences $\dot\beta_i - \dot\beta_k$ tend to nonzero constants as
     $u\to \infty$ [contrary to (\ref{iso})] but $\dot\beta_i$ themselves
     grow infinitely.

     The condition $\Sigma^2 \to 0$ is still weaker. Suppose, for instance,
     that $\beta_2 = \beta_3 \sim \const\cdot u$ as $u\to \infty$ but
     $\beta_1 = \beta_2 + hu$, $h=\const\ne 0$. Then $\cA$ has a nonzero
     limit as $u\to \infty$ whereas $\Sigma^2 = \thd \e^{-2\alpha} h^2 \to
     0$ since the volume factor $v = \e^\alpha \to \infty$.

     Thus {\sl our isotropization condition (\ref{iso}) is stronger than
     the requirements $\cA \to 0$ and $\Sigma^2 \to 0$}.

\subsection{Isotropization conditions}

     With (\ref{sol_23}) and (\ref{int}), the condition (\ref{iso}) for
     $\beta_2$ and $\beta_3$ is fulfilled if and only if
\beq
       c_2 = c_3 = N > 0.                                     \label{N}
\eeq
     It then follows that $\beta_2 \equiv \beta_3$, i.e., the Bianchi
     type I model is plane-symmetric. According to (\ref{sol_23}) and
     (\ref{iso}), $\beta_1 - \beta_2 = 2\beta_1 -Nu \to \const$. Therefore
     at large $u$ we have
\beq                                                          \label{u^}
     \beta\equiv\beta_1 \sim\beta_2 \sim \beta_3 \sim \half Nu; \cm
     		v = \e^\alpha \sim \e^{3Nu/2}.
\eeq
     The physical time is $\tau = \int v(u)\, du \sim \e^{3Nu/2}$, and so the
     possible isotropic expansion of the Universe at large $u$ occurs
     according to the law
\beq                                                         \label{a-iso}
	a_i (\tau) \sim a(\tau) \sim \e^{Nu/2} \sim \tau^{1/3}
		\cm {\rm as} \quad \tau \to \infty.
\eeq

     The asymptotic isotropy condition leads to essential restrictions on
     the properties of the electromagnetic field and the coupling function
     $\Psi$. Indeed, since due to (\ref{u^}) we have $\ddot\beta \to 0$ at
     large $u$, \eq (\ref{dd-beta}) can only be fulfilled if $q_m^2 \Psi +
     q_e^2/\Psi$ tends to zero. This is evidently impossible if both $q_e$
     and $q_m$ are nonzero. We have the following important restriction:

\Theorem{Statement 1}
     {Isotropization is impossible in our system if there are both
     electric and magnetic fields.}

     In what follows, we restrict ourselves to the case of a purely
     magnetic field, $q_m\ne 0$, $q_e=0$. (A reformulation to the case of a
     nonzero electric and zero magnetic field is made by replacing $q_m
     \leftrightarrow q_e$, $\Psi \mapsto 1/\Psi$.)

     Under this assumption, let us obtain a necessary condition for
     isotropization in terms of the coupling function $\Psi$. As follows
     from (\ref{u^}), $\ddot \beta$ vanishes at large $u$, hence
     $\Psi \e^{2\beta}\to 0$, or $\Psi = o(\e^{-Nu})$. Furthermore,
     according to (\ref{int}), $|\dot\phi| \to \pm N\sqrt{3}/2$,
     so that $Nu \approx 2|\phi|/\sqrt{3}$, and we obtain
\beq                                                        \label{Psi-iso}
     \Psi(\phi) = o \Bigl(\e^{- 2|\phi|/\sqrt{3}}\Bigr)
			    	\cm {\rm as} \quad \phi \to \infty.
\eeq
     Under this condition, one can hope to obtain an asymptotically
     isotropic solution by properly fixing the integration constants.

     A physical meaning of the requirement (\ref{Psi-iso}) is that the SET
     components $\Te\mN$ decay more rapidly than $\Ts\mN$, and the model
     becomes scalar field dominated. This agrees with the evolution law
     (\ref{a-iso}), which corresponds to the ultrastiff equation of state $p
     = \rho$ (pressure is equal to energy density), characteristic of a
     massless, minimally coupled, time-dependent scalar field.

     The exponential coupling $\Psi = \e^{2\lambda\phi}$ conforms to
     (\ref{Psi-iso}) provided $\lambda > 1/\sqrt{3}$ in case $\phi\to
     -\infty$ or $\lambda < -1/\sqrt{3}$ in case $\phi\to +\infty$.  Note
     that the ``string'' value $\lambda = \pm 1$ is also admissible.  We
     shall verify these inferences with the exact solution in \sect 4.

\subsection {The singularity problem}

     Let us discuss the possibility of avoiding a cosmological singularity
     in our system without assuming its isotropy at large $u$.

     There is no singularity at finite $u$ since all $|\dot\beta_i| <
     \infty$ and $|\dot\phi| < \infty$. A singularity can occur only
     at $u =\pm\infty$, which, however, can correspond to finite or infinite
     physical time $\tau$ depending on whether the integral
\beq
       \tau = \int \e^{\alpha(u)} du                            \label{tau}
\eeq
     converges or diverges. We can assert the following:

\Theorem{Statement 2}
    {For the models under study, the range of $\tau$ is either
     $-\infty < \tau < \tau_0$ or $\tau_0 < \tau <+\infty$ where $\tau_0$
     is finite.  The volume factor $v=\e^\alpha$ is a strictly monotonic
     function of $u$.  }

     Indeed, it follows from (\ref{int}) that
\beq
	\dot \beta^2 \leq c_2 c_3,                              \label{b<}
\eeq
     so, in a nontrivial solution, the constants $c_2$ and $c_3$ should be
     nonzero and have the same sign. Suppose that they are positive, then
     the evident chain of inequalities
\[
        c_2 c_3 < 2 c_2 c_3 < (c_2 + c_3)^2
\]
     implies that the quantity $\dot\alpha = -\dot \beta + c_2 + c_3$ is
     strictly positive (and finite) at all $u$, including the limits
     $u \to \pm\infty$, i.e., $\alpha(u)$ monotonically grows. The integral
     (\ref{tau}) thus converges at $u\to -\infty$ and diverges at $u
     \to+\infty$, which means that $\tau_0 < \tau <+\infty$. If $c_2$ and
     $c_3$ are negative, we have a strictly monotonically decreasing
     function $\alpha(u)$ and $-\infty < \tau < \tau_0$. A specific model
     can be either eternally expanding (with $dv/d\tau > 0$) or eternally
     contracting ($dv/d\tau <0$).

     Suppose, without loss of generality, $c_2 > 0$ and $c_3 >0$, thus
     choosing an expanding model. As $u\to \infty$, $\tau \sim
     \e^{\const\cdot u}$, while the functions $|\beta_i (u)|$ also grow at
     most exponentially, therefore the scale factors $a_i (\tau)$ cannot
     grow or vanish faster than according to a power law. This means (see
     the Appendix) that the model is nonsingular at the end of the
     evolution.

     On the contrary, at its beginning $\tau\to \tau_0$, where $v = a_1 a_2
     a_3 \to 0$, the only way of avoiding a singularity is to assume that
     only one of the scale factors $a_i$ vanishes while the others remain
     finite.  If we suppose that it is $a_2$ or $a_3$ that vanishes, then
     from (\ref{sol_23}) we immediately obtain that one of the constants
     $c_2$, $c_3$ is zero, making the solution trivial. Therefore the only
     viable opportunity is
\beq
      a_1 \to 0, \cm a_2 = a_3 \to \const >0,                 \label{as-1}
\eeq
     which happens when $c_2 = c_3 = N > 0$ (as was the case for
     isotropization). Assigning $\tau_0 =0$, we obtain $a_1 \sim \tau \sim
     \e^{Nu} \to 0$ as $u\to -\infty$, and the metric near $\tau = 0$ may be
     written as
\beq                                                          \label{ds-1}
      ds^2 \approx d\tau^2 - \frac{k_1^2}{\tau^2} {dx^1}^2
      			   - k_2^2({dx^2}^2 + {dx^3}^2)
		 = \frac{k_1}{2t}\,dt^2 - \frac{2t}{k_1}\, {dx^1}^2
      			   - k_2^2({dx^2}^2 + {dx^3}^2)
\eeq
     where $k_1,\ k_2 = \const > 0$ and $ t = \tau^2/(2k_1)$. Evidently,
     the instant $\tau = t = 0$ is a Killing horizon, and a transition to
     negative $t$ leads to a static, plane-symmetric space-time region where
     $t$ is a spatial coordinate and $x^1$ temporal. The properties of the
     region $t < 0$ may be studied for specific $\Psi(\phi)$.

     \eq (\ref{int}) shows that at such a horizon $\dot\phi\to 0$
     and $\Psi \e^{2\beta} \to 0$. Finiteness of $\Ts\mN$ implies that
     $\dot\phi$ vanishes as $\e^{Nu}=\e^{-N|u|}$ or faster. On the other
     hand, the magnetic field strength is finite at the horizon [see
     (\ref{E,B})], and finiteness of $\Te\mN$ implies $\Psi < \infty$.

     One can also note that since both isotropization and a horizon in
     the past require $c_2 = c_3$, therefore one cannot exclude that some
     solutions possess both features.

     The main conclusion from this general study can be formulated as
     follows.

\Theorem{Statement 3}
    {Expanding models begin their evolution at finite proper time $\tau$
     (say, $\tau=0$) either from a singularity, which is the general case,
     or from a simple horizon preceded by a static phase. The latter happens
     when the model possesses additional planar symmetry [$a_2(\tau)
     \equiv a_3(\tau)$] and, moreover, both $a_2$ and $a_3$ take a finite
     value at $\tau=0$.  }

     A reformulation for contracting models is obvious.

\section{Exact solution for $\Psi = \e^{2\lambda\phi}$}

\subsection {Solution}

     We choose the function $\Psi$ in the form $\Psi = \e^{2\lambda \phi}$,
     $\lambda=\const$, returning to the action (\ref{dil}).
     Our set of equations (\ref{dd-phi})--(\ref{int}) is then rewritten as
\bearr
	\ddot \phi = -\lambda q_m^2 \e^{2\beta + 2\lambda\phi},\label{dd-fi}
\\ \lal
	\ddot \beta = -q_m^2 \e^{2\beta + 2\lambda\phi},       \label{dd-b}
\\ \lal                                                        \label{int'}
	\dot\beta{}^2
	 	+ \dot\phi{}^2 + q_m^2 \e^{2\beta + 2\lambda\phi} = c_2 c_3.
\ear
     \eqs (\ref{dd-fi}) and (\ref{dd-b}) lead to
\bearr                                                       \label{la-fi}
	\ddot\phi = \lambda\ddot\beta  \then  \phi = \lambda\beta-c_1 u,
\\ \lal
	\ddot y = -q_m^2(1+\lambda^2)\e^{2y}                   \label{y}
			     \ \ \then \ \
		\e^y = \frac{k}{q_m\sqrt{1+\lambda^2} \cosh (ku)},
\ear
     where $y(u) := \beta + \lambda\phi$, $c_1$ and $k >0 $ are constants.
     Two more integration constants are absorbed by rescaling $x^1$ and by
     choosing the origin of $u$. The functions $\phi$ and $\beta$ are
     expressed in terms of $y$:
\beq
	\phi = \frac{\lambda y -c_1 u}{1+\lambda^2},   \cm
	\beta= \frac{y + \lambda c_1 u}{1+\lambda^2},            \label{sol}
\eeq
     The constraint (\ref{int'}) leads to a relation between the integration
     constants:
\beq
     k^2 + c_1^2 = (1+\lambda^2) c_2 c_3.                      \label{int+}
\eeq
     This, along with \eq (\ref{sol_23}), completes the solution. Apart from
     the coupling constant $\lambda$, the general solution contains five
     integration constants: $q_m,\ k,\, c_1,\ c_2,\ c_3$ with the single
     constraint \eq (\ref{int+}).

     An explicit form of the solution is
\bear
     a_1 (u) \eql \biggl[
     	  \frac{\e^{\lambda c_1 u}}{Q\cosh(ku)}\biggr]^{1/(1 + \lambda^2)},
\nn
     a_2 (u) \eql \e^{c_2 u} \biggl[
       \frac{\e^{\lambda c_1 u}}{Q\cosh(ku)}\biggr]^{-1/(1 + \lambda^2)},
\cm
     a_3 (u) = \e^{c_3 u} \biggl[
       \frac{\e^{\lambda c_1 u}}{Q\cosh(ku)}\biggr]^{-1/(1 + \lambda^2)},
\nn
     \e^\phi \eql [Q\cosh (ku)]^{-\lambda/(1 + \lambda^2)}
		    			\e^{-c_1 u/(1 + \lambda^2)},
\nnv
     d\tau \eql a_1(u)\, a_2(u)\, a_3(u)\, du,                 \label{sol-E}
\ear
     where $Q = (q_m/k) \sqrt{1+ \lambda^2}$.

     This solution coincides, up to notations, with that found in
     \cite{baner,harko}.

\subsection {Isotropization}

     The Bianchi-I cosmologies described by \eqs (\ref{sol-E}) begin with
     an anisotropic singularity at finite $\tau$ corresponding to
     $u \to -\infty$ and, in general, expand anisotropically at late times
     ($u\to\infty, \tau\to \infty$).

     Let us, however, seek among them a subfamily of models becoming
     isotropic as $\tau\to +\infty$. As before, we put $c_2 = c_3 = N > 0$;
     one more constraint follows from the requirement $\dot \beta (+\infty)
     = N/2$ (see \sect 3.2) which leads to
\beq
	\lambda c_1 -k = (1 + \lambda^2) N/2.                  \label{iso-2}
\eeq
     Then, as follows from (\ref{int'}),
     $\dot\phi (+\infty) = \pm N \sqrt{3}/2$. We choose, without loss of
     generality, the minus sign, so that $\phi \to -\infty$, which implies
     $\lambda >0$ since we must have $\Psi = \e^{2\lambda\phi} \to 0$ (see
     \sect 3.2). As a result, $c_1$ and $k$ are expressed in terms of $N$
     and $\lambda$:
\beq
	2c_1 = N(\lambda + \sqrt{3}), \cm
	  2k = N(\lambda \sqrt{3} -1).                         \label{iso-3}
\eeq
     We thus obtain a family of asymptotically isotropic solutions
     parametrized by $\lambda$ and two integration constants $N$ and $q_m$.
     \eq (\ref{iso-3}) implies the requirement
\beq
	\lambda > 1/\sqrt{3},                                  \label{lam>}
\eeq
     in full agreement with \eq (\ref{Psi-iso}).

\subsection {The horizon and beyond}

     To find solutions with a horizon instead of a singularity in the remote
     past, we put again $c_2 = c_3 = N > 0$ and also require (see \sect 3.3)
     $\beta_2 = \beta_3 \to \const$ and $\dot \phi \to 0$ as $u\to -\infty$.
     We obtain
\beq
	c_1 = k \lambda, \cm   k = N.                        \label{hor-1}
\eeq
     The first of these conditions also provides a finite value of $\phi$ at
     the horizon.

     We thus obtain one more family of solutions parametrized by $\lambda$,
     $N$ and $q_m$. It is, in general, different from the family of
     asymptotically isotropic solutions and, in particular, it exists for
     any $\lambda$. However, the two families coincide in case
     $\lambda = \sqrt{3}$. {\sl In this and only in this case we have a
     two-parameter family of asymptotically isotropic solutions without a
     cosmological singularity.}

     Let us continue the solutions satisfying (\ref{hor-1}) beyond their
     horizon. To this end, we use the coordinate transformation
\beq
      \e^{2ku} = \xi - 1                                      \label{u->xi}
\eeq
     and introduce the notations
\beq
      \mu := \frac{1}{1 + \lambda^2}, \cm                     \label{C_0}
      C_0 := \biggl(\frac{2k}{q_m \sqrt{1+\lambda^2}}\biggr)^{2\mu}.
\eeq
     As a result, we obtain the metric
\bear                                                         \label{ds-hor}
      ds^2 = \frac{\xi^{2\mu}}{4k^2 C_0}\, \frac{d\xi^2}{\xi-1}
	     - \frac{C_0 (\xi-1)}{\xi^{2\mu}} {dx^1}^2
	     - \frac{\xi^{2\mu}}{C_0} \Bigl({dx^2}^2 + {dx^3}^2\Bigr).
\ear
     The horizon takes place at $\xi =1$, and $\xi > 1$ describes the
     cosmological evolution, which is asymptotically isotropic at large
     $\xi$ if and only if $\mu = 1/4$ which corresponds to $\lambda^2 =3$,
     in full agreement with the above-said.

     In the region $0 < \xi < 1$, the metric (\ref{ds-hor}) is static,
     plane-symmetric, $x^1$ is a temporal coordinate and $\xi$ is the
     spatial coordinate in the direction across the symmetry planes
     parametrized by $x_2$ and $x_3$. Lastly, $\xi = 0$ is a timelike
     repulsive singularity resembling the one in the Reissner-Nordstr\"om
     space-time. It is connected with infinite fields, densities and
     stresses: the magnetic field strength $B$ and both $\Te\mN$ and
     $\Ts\mN$ are infinite at $\xi =0$.

\section {String metric}

     In case $\lambda = 1$, the Einstein-frame action (\ref{dil}) with the
     metric $g\mn$ is obtained by the conformal transformation
\beq
       \hg\mn = \e^{-2\phi} g\mn                             \label{trans}
\eeq
     from the string-frame action
\beq
      S = \int \sqrt{\hg} d^4x \e^{2\phi }             		\label{S}
      			\Bigl[ \hat R
		      - 4 (\hat{\d}\phi)^2 - {\hat F}\MN F\mn \Bigr],
\eeq
     where the hat marks quantities obtained from or with the aid of the
     so-called string metric $\hg\mn$. The action in the form (\ref{S})
     naturally appears from string theory, and in this sense the metric
     $\hg\mn$ is more fundamental than $g\mn$, although the question ``which
     metric is to be used to confront theory with observations?'' should be
     answered separately --- see more detailed discussions in
     Refs.\,\cite{frames,br95} and references therein.

     Now, let us take the coupling constant $\lambda = 1$ and briefly
     discuss the properties of the metric $\hg\mn$ corresponding to the
     solution of \sect 4. Evidently, $\ha_i (u) = \e^{-\phi} a_i(u)$, and the
     proper time element is $d\hat \tau = \e^{-\phi} d\tau$. Explicitly,
     using (\ref{sol-E}) with $\lambda=1$, it is straightforward to obtain
\bear
      \e^\phi \eql \e^{-c_1 u/2} [Q \cosh (ku)]^{-1/2},     \label{phi-S}
\nn
     d\hat s^2 \eql d\hat\tau^2 - \sumi \ha_i^2 (dx^i)^2
\nn
     \eql Q^2 \e^{2(c_2 + c_3)u} \cosh^2 (ku) du^2           \label{ds-S}
	  - \e^{2c_1u} {dx^1}^2
	- Q^2 \cosh^2 (ku)\Bigl(\e^{2c_2u}{dx^2}^2+\e^{2c_3u}{dx^3}^2\Bigr),
\ear
     where $q_m \sqrt{2}/k$. Choosing, as before, positive $c_2$ and $c_3$,
     we see that in expanding models $\hat \tau \to \infty$ corresponds to
     $u\to \infty$ while $u \to -\infty$ corresponds to finite $\hat \tau$.

     Furthermore, the asymptotic isotropy conditions are the same as in the
     Einstein frame since conformal mappings do not affect the condition
     (\ref{iso}). Specialized to $\lambda=1$, they read
\beq
       c_2 = c_3 = N, \cm
      2c_1 = N(\sqrt{3} + 1), \cm
	 k = N(\sqrt{3} -1).
\eeq

     One may notice that, in the Einstein frame, the expansion begins from a
     Killing horizon instead of a singularity in case $\phi\to\const$ as
     $u \to -\infty$, i.e., the conformal factor in (\ref{trans}) is finite
     at the horizon. Therefore the conditions (\ref{hor-1}) preserve their
     meaning in the string frame and now read
\beq
	c_1 = k = N. 				            \label{hor-2}
\eeq
     After the substitution (\ref{u->xi}), the metric (\ref{ds-S}) with
     (\ref{hor-2}) transforms to
\beq                                                        \label{S-hor}
     d\hat s^2 = \frac{Q^2}{4k} \frac{\xi^2}{\xi-1} d\xi^2
	     - (\xi-1) {dx^1}^2 - \frac{Q^2}{4} \xi^2
    	                                  \Bigl({dx^2}^2+{dx^3}^2\Bigr).
\eeq
     As in (\ref{ds-hor}), the cosmological evolution corresponds to $u\in
     \R$, or $\xi > 1$, while $0 < \xi < 1$ is a static region with a
     singularity at $\xi = 0$. This singularity is, in a sense, milder than
     in the Einstein frame since it is purely spatial: the temporal metric
     coefficient $\hg_{11}$ is finite at $\xi = 0$ (recall that it is $x^1$
     that plays the role of time at $\xi < 1$).

\section {Concluding remarks}

     We have described some important qualitative features of Bianchi-I
     cosmologies in the theory (\ref{S-Psi}), containing the arbitrary
     function $\Psi(\phi)$, without entirely solving the field equations ---
     see items 1--5 in the Introduction. These features have been confirmed
     in an exactly solvable particular case, the system (\ref{dil}).

     For this latter system, exact solutions were already obtained in
     Refs.\,[2--4]; according to \cite{baner}, the solutions given in
     \cite{salim} contained an error; those given in \cite{baner,harko}
     coincide with ours up to notations. However, the authors of these
     papers did not point out any isotropizing models or models whose
     expansion starts from a horizon.

     We have shown that for this system, asymptotically isotropic solutions
     form a two-parameter family for each suitable $\lambda$ whereas the
     full set contains four essential integration constants. In this sense,
     isotropization requires fine tuning. The same is true for solutions
     having a horizon instead of a singularity at the beginning of their
     expansion. For a particular value of the coupling constant $\lambda$
     ($\pm \sqrt{3}$) these two-parameter families coincide.

     Other examples of cosmologies whose expansion starts from a horizon
     (from the so-called ``null bang'') were suggested in
     Ref.\,\cite{br-dym03} as vacuum cosmologies with a variable
     cosmological term; unlike the present examples, they contained a de
     Sitter-like core instead of a singularity in their static region.

     Certain remarks are to be made about other exact solutions to the
     Einstein-Maxwell-dilaton equations due to (\ref{dil}).

     First of all, some solutions can be obtained with both electric and
     magnetic charges. \eqs (\ref{dd-phi}) and (\ref{dd-beta}) with $\Psi =
     \e^{2\lambda\phi}$ read
\beq                                                          \label{e-dub}
      \ddot\phi = -\lambda q_m^2 \e^{2y} + \lambda q_e^2 \e^{2z},
\cm
      \ddot\beta = -q_m^2 \e^{2y} - q_e^2 \e^{2z},
\eeq
     where $y = \beta + \lambda\phi$ and $z = \beta - \lambda\phi$. Their
     two linear combinations form the Toda-like system
\bear                                                         \label{toda}
     \ddot y \eql -q_m^2 (1+\lambda^2) \e^{2y} - q_e^2 (1-\lambda^2)\e^{2z},
\nn
     \ddot z \eql -q_m^2 (1-\lambda^2) \e^{2y} - q_e^2 (1+\lambda^2)\e^{2z}.
\ear
     These equations decouple and take an easily solvable Liouville form
     [see (\ref{y})] in case $\lambda^2 = 1$, i.e., for the string value of
     the electro-dilatonic coupling constant. We will not proceed with this
     solution, only recalling that it evidently cannot describe isotopizing
     cosmologies --- see Statement 1 above.

     The distinguished properties of string coupling, leading to
     integrability of the Einstein-Maxwell-dilaton equations, were
     previously discussed for multidimensional \ssph\ configurations with
     the Lagrangian (\ref{dil}) in Ref.\,\cite{br95} where solutions with
     three charges of different nature were obtained (in addition to electric
     and magnetic charges, there is a quasiscalar charge related to possible
     extra-dimensional components of $F\mn$), including as special cases
     different dilatonic black holes. Our present solutions (\ref{sol}),
     (\ref{sol-E}) with a single charge have a deep similarity with \ssph\
     solutions for the system (\ref{dil}) in Ref.\,\cite{bsh77} where also,
     in a special case, a horizon was found instead of a naked singularity
     (to our knowledge, that was the first example of what was later named a
     dilatonic black hole).

     Still closer counterparts of the present solutions are static,
     plane-symmetric and cylindrically symmetric ones (see, e.g.,
     \cite{br79} for the corresponding Einstein-Maxwell solutions, including
     a special case with a horizon).

     In a more general context, the system (\ref{dil}) is a special case of
     systems with self-gravitating interacting scalars and antisymmetric
     forms in diverse dimensions, frequently associated with $p$-branes and
     appearing in the bosonic sector of the field limit of supergravities,
     M-theory, etc (see \cite{M-th} and references therein). Methods of
     obtaining exact solutions for large classes of such models have been
     elaborated (\cite{pbra-sol} and references therein) on the basis of a
     sigma model representation where each charge like $q_m$ and $q_e$ is
     associated with a constant vector in a target space. In terms of this
     approach, the case $\lambda^2 = 1$ of \eqs (\ref{toda}) corresponds to
     mutual orthogonality of two such vectors; for other values of $\lambda$
     the orthogonality is lacking, and the equations cannot be so easily
     integrated.

\section*{Appendix}
\def\theequation{A.\arabic{equation}}\sequ{0}

     Consider regularity conditions for the metric (\ref{B1}) with the
     aid of the Kretschmann scalar $\cK = R_{\alpha\beta\gamma\delta}
     R^{\alpha\beta\gamma\delta}$. As in many other cases, $\cK$ for this
     metric is a sum of squares of all nonzero Riemann tensor components
     $R_{\alpha\beta}{}^{\gamma\delta}$. Therefore the condition $\cK <
     \infty$ is necessary and sufficient for finiteness of all algebraic
     curvature invariants. Explicitly, for (\ref{B1}) written with an
     arbitrary $u$ coordinate,
\beq                                                         \label{K-gen}
     \cK = 4 \sumi
	\left[\e^{-\alpha-\beta_i} (\e^{\beta_i - \alpha}\dot\beta_i)
		\dot{\tst\mathstrut}\ \right]^2
	+ 4 \sum_{i\ne k}
		\left[\e^{-2\alpha}\dot\beta_i\dot\beta_k\right]^2.
\eeq
     In terms of the scale factors $a_i = \e^{\beta_i}$ as functions of the
     physical time $\tau$,
\beq                                                         \label{K-phys}
     \cK = 4 \sumi \biggl(\frac{a''_i}{a_i}\biggr)^2
	 + 4 \sum_{i\ne k}
	 	\biggl(\frac{a'_i}{a_i}\,\frac{a'_k}{a_k}\biggr)^2
\eeq
     where the prime stands for $d/d\tau$. This expression leads to evident
     and convenient regularity criteria.

     Our solutions, written in terms of the harmonic time coordinate $u$,
     are manifestly regular at finite $u$. Meanwhile, $u \to \pm\infty$ may
     correspond to both finite and infinite physical time $\tau$. Let us
     therefore discuss the regularity condition $\cK < \infty$ at finite and
     infinite $\tau$ separately. Evidently, a singularity can occur when
     some or all scale factors $a_i$ tend to zero or infinity.

\begin{description}
\item[(i)] $\tau$ finite, some $a_i\to \infty$. Since in this case
     $\ln a_i$, $a'_i$ and $\ln a'_i$ blow up as well, we have $a''_i/a_i =
     (a''_i/a'_i)(a'_i/a_i)\to \infty$. Therefore, {\sl if at least
     one scale factor becomes infinite at finite $\tau$, it is a curvature
     singularity.}

\item[(ii)] $\tau$ finite, some $a_i\to 0$. Since $\ln a_i \to -\infty$,
     we have $a'_i/a_i \to -\infty$. However, $a''_i/a_i$ may remain finite.
     The expression (\ref{K-phys}) shows that {\sl if more than one scale
     factor turns to zero at finite $\tau$, it is a singularity. If only one
     scale factor $a_i = 0$ at finite $\tau$, the space-time can be
     nonsingular.} An explicit inspection is then necessary: one may find,
     e.g., a Killing horizon instead of a singularity.

\item[(iii)] $\tau \to \infty$, some $a_i \to \infty$. An inspection
     shows that {\sl such an asymptotic can only be singular if at least one
     function $a_i(\tau)$ grows faster than exponentially:}
     $a_i (\tau) \gg \exp(k |\tau|)$, $k = \const >0$.

\item[(iv)] $\tau \to \infty$, some $a_i \to 0$. As in item (iii), we find
     that {\sl such an asymptotic can only be singular if at least one
     function $a_i(\tau)$ vanishes faster than exponentially:}
     $a_i (\tau) = o (\exp[-k |\tau|])$, $k = \const >0$.
\end{description}

\subsection*{Acknowledgment}

We acknowledge partial financial support from the Ministry of Education of
Russia.

\end{document}